\documentclass[sigconf]{acmart}

\AtBeginDocument{%
  }
\usepackage{algorithm}
\usepackage{algpseudocode}

\copyrightyear{2026}
\acmYear{2026}
\setcopyright{cc}
\setcctype{by}
\acmConference[PQQS '26]{Post-Quantum and Quantum-based Security}{November 02--04, 2026}{Milpitas, CA, USA}
\acmBooktitle{Post-Quantum and Quantum-based Security (PQQS '26), November 02--04, 2026, Milpitas, CA, USA}
\acmDOI{10.1145/3839338.3857253}
\acmISBN{979-8-4007-2920-1/2026/11}

\graphicspath{%
  {sections/images/}%
  {sections/images/training/}%
  {sections/images/controller/}%
  {sections/images/methodology/}%
}

\newcommand{\Fmin}{F_{\min}}
\newcommand{\Freq}{F_{\mathrm{req}}}
\newcommand{\Fdel}{\bar{F}_{\mathrm{del}}}
\newcommand{\Fetoe}{F_{\mathrm{e2e}}}
\newcommand{\Tc}{T_c}
\newcommand{\Text}{T_c^{\mathrm{ext}}}
\newcommand{\tcut}{t_{\mathrm{cut}}}
\newcommand{\uskr}{u_{\mathrm{skr}}}
\newcommand{\pgen}{p_{\mathrm{gen}}}
\newcommand{\Latt}{L_{\mathrm{att}}}
\newcommand{\tick}{\tau}

\begin{document}

\title{Sequential vs. Synchronized Entanglement Swapping: QKD Secret-Key Rate under Optimized Link-Layer Control}

%% ---- Authors (in the order supplied for ACM metadata) ------------
\author{Priyam Srivastava}
\email{PRS216@pitt.edu}
\affiliation{%
  \department{Department of Informatics and Networked Systems}
  \institution{University of Pittsburgh}
  \city{Pittsburgh}
  \state{Pennsylvania}
  \country{USA}
}

\author{Akshat R. Sabavat}
\email{ARS789@pitt.edu}
\affiliation{%
  \department{Department of Information Science}
  \institution{University of Pittsburgh}
  \city{Pittsburgh}
  \state{Pennsylvania}
  \country{USA}
}

\author{Siddharth Jain}
\email{SIJ50@pitt.edu}
\affiliation{%
  \department{Department of Physics and Astronomy}
  \institution{University of Pittsburgh}
  \city{Pittsburgh}
  \state{Pennsylvania}
  \country{USA}
}

\author{Alan Scheller-Wolf}
\email{awolf@andrew.cmu.edu}
\affiliation{%
  \department{Tepper School of Business}
  \institution{Carnegie Mellon University}
  \city{Pittsburgh}
  \state{Pennsylvania}
  \country{USA}
}

\author{Sridhar Tayur}
\email{stayur@andrew.cmu.edu}
\affiliation{%
  \department{Tepper School of Business}
  \institution{Carnegie Mellon University}
  \city{Pittsburgh}
  \state{Pennsylvania}
  \country{USA}
}

\author{David Tipper}
\email{tipperdavid@gmail.com}
\affiliation{%
  \department{Department of Informatics and Networked Systems}
  \institution{University of Pittsburgh}
  \city{Pittsburgh}
  \state{Pennsylvania}
  \country{USA}
}

\author{Prashant Krishnamurthy}
\email{prashk@pitt.edu}
\affiliation{%
  \department{Department of Informatics and Networked Systems}
  \institution{University of Pittsburgh}
  \city{Pittsburgh}
  \state{Pennsylvania}
  \country{USA}
}

\author{Amy Babay}
\email{babay@pitt.edu}
\affiliation{%
  \department{Department of Informatics and Networked Systems}
  \institution{University of Pittsburgh}
  \city{Pittsburgh}
  \state{Pennsylvania}
  \country{USA}
}

\author{Kaushik Seshadreesan}
\email{kausesh@pitt.edu}
\affiliation{%
  \department{Department of Informatics and Networked Systems}
  \institution{University of Pittsburgh}
  \city{Pittsburgh}
  \state{Pennsylvania}
  \country{USA}
}

\renewcommand{\shortauthors}{Srivastava et al.}

%% ---- Abstract ----------------------------------------------------

\begin{abstract}
Quantum repeater chains build long-distance entanglement for quantum key
distribution (QKD) by creating pairs on short links and joining them through
entanglement swapping. Because stored pairs lose fidelity while they wait,
swap timing can affect the secret-key rate. We study how network-layer
swap scheduling affects six-state QKD on a simulated four-hop chain under
an idealized Werner-state noise model. We compare a sequential
(swap-and-wait) discipline, which builds and stores partial chains,
with a synchronized (wait-and-swap) discipline, which waits until all
links are ready before swapping. Both use the same reinforcement-learned
link controllers, isolating the effect of the swapping discipline.
When stored pairs cannot retain sufficient fidelity for one network
time step, our sequential implementation produces no key, while
synchronized retains a small key rate from simultaneous link deliveries.
Once sequential becomes viable, it outperforms synchronized by up to
$19.1\%$ in the homogeneous chains studied, with the advantage
decreasing as memory lifetime increases. We predict the synchronized
short-memory pair-delivery rate from link-delivery statistics and
derive a closed-form condition for when sequential can preserve
progress across time steps. We also examine chains with one slower
link, where partial chains can wait and decay at the bottleneck.
In this setting, synchronized slightly outperforms sequential at
longer memory lifetimes. Moving the slow link has little effect on
key rate away from the transition region but strongly affects
sequential link-buffer occupancy. Choosing a swapping discipline
therefore depends not only on whether memories can preserve partial
progress, but also on how the link configuration determines where
that progress must wait and decay.
\end{abstract}

%% ---- CCS + keywords ----------------------------------------------
\begin{CCSXML}
<ccs2012>
   <concept>
       <concept_id>10002978.10002979</concept_id>
       <concept_desc>Security and privacy~Cryptography</concept_desc>
       <concept_significance>500</concept_significance>
       </concept>
   <concept>
       <concept_id>10003033.10003039.10003045</concept_id>
       <concept_desc>Networks~Network layer protocols</concept_desc>
       <concept_significance>500</concept_significance>
       </concept>
   <concept>
       <concept_id>10003033.10003079.10003081</concept_id>
       <concept_desc>Networks~Network simulations</concept_desc>
       <concept_significance>500</concept_significance>
       </concept>
 </ccs2012>
\end{CCSXML}

\ccsdesc[500]{Security and privacy~Cryptography}
\ccsdesc[500]{Networks~Network layer protocols}
\ccsdesc[500]{Networks~Network simulations}

\keywords{quantum networks, quantum key distribution, secret-key rate,
  entanglement distribution, entanglement swapping, swap scheduling,
  quantum repeaters, packet switching, memory decoherence,
  network-layer control, reinforcement learning}

\maketitle

%% ---- Body: one \input per section --------------------------------

\section{Introduction}
\label{sec:intro}

Quantum networks share entanglement between distant nodes to support
applications that classical networks cannot~\cite{wehner2018quantum}.
These include distributed quantum computing~\cite{cirac1999distributed,CALEFFI2024110672},
quantum-enhanced sensing~\cite{komar2014quantum}, and entanglement-based
quantum key distribution (QKD)~\cite{ekert1991quantum}.
Over long distances, a repeater chain first creates entangled pairs
over short hops between neighboring nodes, called \emph{elementary links},
and then joins them into an end-to-end pair through \emph{entanglement
swapping}~\cite{briegel1998repeaters}. While pairs wait in quantum memory
to be swapped, they decohere and lose fidelity. The secret-key rate
therefore depends not only on how quickly the links create entanglement,
but also on when the network performs swaps and how long intermediate
entanglement is stored. Because QKD produces key only when errors remain
sufficiently low, the swap schedule can affect both the rate at which a chain
produces key and whether it produces any key at all.

Swap scheduling in repeater chains has largely been studied with delivery
time as the objective. I\~{n}esta et al.~\cite{inesta2023optimal} model a
repeater chain as a Markov decision process and compute policies that
minimize the expected time to deliver an end-to-end pair under memory
cutoffs. Their baseline, swap-as-soon-as-possible (SWAP-ASAP), allows a
node to swap as soon as both adjacent links are
ready~\cite{inesta2023optimal,coopmans2022improved}. Other work analyzes
sequential, parallel, and asynchronous entanglement-distribution
strategies under simplified link
models~\cite{deandrade2024sequential,pouryousef2024analysis}. For QKD,
however, delivery time alone is not sufficient: the key rate
depends on both how often end-to-end pairs arrive and how much fidelity
they retain. This motivates evaluating network control using an
application-level utility such as secret-key
rate~\cite{vardoyan2023qnum, li_pnas2022}.

A complementary line of work studies how the elementary links themselves
should be controlled. Yau et al.~\cite{yau2025rl} use reinforcement
learning to train link controllers for nonlinear application objectives
such as secret-key rate. Reinforcement learning has also been used to find
cutoff and collaboration policies in repeater
chains~\cite{haldar2024fast}. Other work considers the classical
information available to network controllers, including quasi-local
policies~\cite{haldar2025reducing}, communication
delays~\cite{li2024commdelay}, and optimistic operation before heralding
messages arrive~\cite{mobayenjarihani2024optimistic}. Together, these
lines of work show that both link-level control and network-layer
scheduling affect entanglement distribution, but leave open how different
swap schedules compare under the same QKD-optimized link layer.

We study this question by comparing two contrasting network-layer swapping
disciplines. \emph{Sequential} (swap-and-wait) builds an end-to-end pair
one hop at a time, storing a partial chain until the next link is ready.
This is akin to the packet-switched design of Bacciottini
et al.~\cite{bacciottini2025internet,bacciottini2025packet}, in which
entanglement is forwarded hop by hop without reserving an end-to-end route
in advance. \emph{Synchronized} (wait-and-swap) instead waits until every
link holds a usable pair and then joins the full chain at once.
This resembles circuit-switched operation in that resources along the
end-to-end path must be available before the swapping operation proceeds.
Both disciplines may store elementary-link pairs, but only sequential
stores partially completed chains. They therefore represent two extremes
in how much partial progress the network preserves. Strategies between
these extremes may outperform either one, and we do not study them here.

To isolate the effect of this network-layer choice, each elementary link
uses the same reinforcement-learned controller under both disciplines.
The controller uses a Werner-state model of an elementary link with
two nodes and two quantum memories per node, referred to as WN2M2 by
Yau et al.~\cite{yau2025rl}, and is trained to maximize the secret-key
rate that the link alone would support under six-state
QKD~\cite{bruss1998sixstate}. It may wait, discard, purify, or deliver an
entangled pair to the network layer. Each delivered pair must exceed a
minimum fidelity $\Freq(n)$ chosen so that $n$ such pairs can still
produce key after swapping (Section~\ref{sec:methods-budget}).
The link controller does not observe the network-layer discipline, so
differences between sequential and synchronized arise from how the
same link-layer deliveries are stored and swapped.

We evaluate a four-hop chain ($n=4$), the chain length also used to
evaluate the packet-switched design~\cite{bacciottini2025internet}.
One entanglement-generation attempt on a link of length $L$ takes
$\tau=L/c$. For homogeneous chains, this also sets the network time
step, and we vary memory coherence relative to $\tau$ through
$\Tc/\tau$. We study two link lengths, first in homogeneous chains and
then in heterogeneous chains containing one slower link at different
positions. In the heterogeneous case, the network advances in steps
of the shortest link's attempt period, $\tau_{\min}$. A link whose
attempt period is $k\tau_{\min}$ is scheduled once every $k$ network
time steps, where $k$ is an integer.
Generation time is included, while other messaging delays and local
operations are treated as instantaneous. We consider one fixed path
with no routing or competing users. The study therefore compares two
swapping disciplines under a common link layer rather than two
complete network architectures.

We make three contributions.
\begin{itemize}
\item \textbf{A network-layer comparison of sequential and synchronized
swapping.}
When stored entanglement cannot remain usable across network time steps,
the sequential implementation produces no key, while synchronized
retains a low key rate. Once sequential becomes viable, it leads in
homogeneous chains by up to $19.1\%$ at $L=5$~km and $7.0\%$ at
$L=10$~km, with the advantage falling below $1\%$ at the longest
memory lifetimes. With one slower link, however, synchronized slightly
outperforms sequential at longer memory lifetimes. The bottleneck
position has little effect on key rate away from the transition but
strongly affects sequential link-buffer occupancy.

\item \textbf{Analytical explanations for the short-memory regime and
sequential viability.}
When stored pairs cannot remain usable for one network time step,
synchronized can still succeed when all links deliver together.
We predict this short-memory pair-delivery rate from link-delivery
statistics without fitting. We also derive a closed-form condition
on $\Tc/\tau$ for when a delivered pair can survive long enough
for sequential operation. For our controllers, it gives thresholds
of approximately $142$--$155$, consistent with sequential producing
no key at $\Tc/\tau=100$ and producing key at $\Tc/\tau=250$.

\item \textbf{QKD-oriented reinforcement learning for the link layer.}
Below the fidelity needed to generate key, the QKD objective is flat
at zero and provides no useful training signal, which can cause the
controller to continue purifying without reaching a useful policy.
We introduce a bootstrap training rule that rewards fidelity while
penalizing excessive delivery time until the key-producing region is
reached. Across the $24$ trained controllers used in our evaluation,
the average delivered fidelity exceeds the four-hop requirement
of $0.9476$.
\end{itemize}

The rest of the paper is organized as follows.
Section~\ref{sec:methods} presents the model and swapping disciplines;
Section~\ref{sec:results} reports the simulation results; and
Section~\ref{sec:discussion} provides analytical explanations and
discusses limitations and extensions.
Section~\ref{sec:conclusion} concludes the paper.

\section{Methodology}
\label{sec:methods}

\begin{figure*}[htbp]
  \centering
  \includegraphics[width=\textwidth]{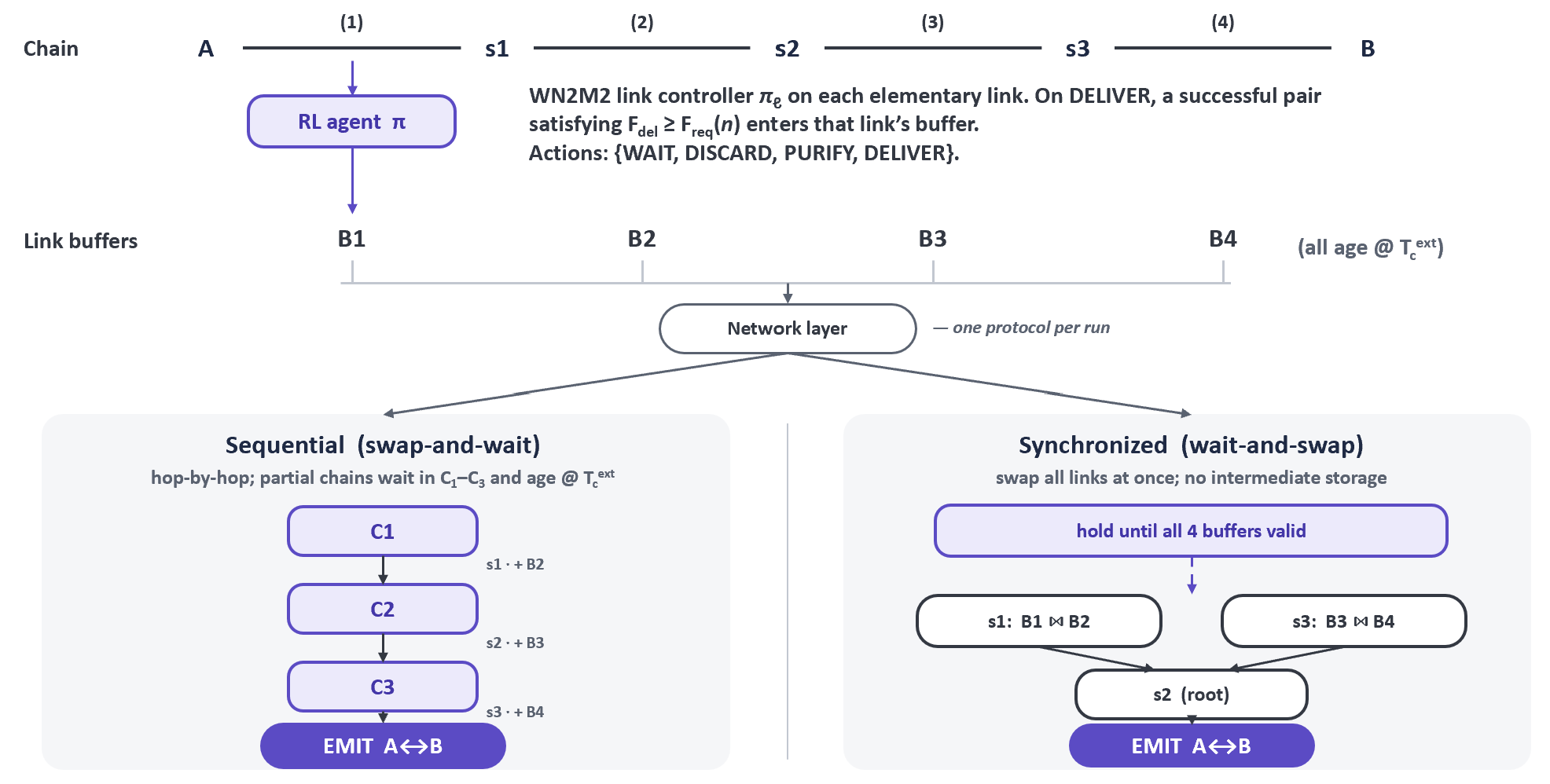}
  \caption{Two-layer architecture for an $n=4$ chain. Top: each
    elementary link runs a WN2M2 controller with actions
    \{WAIT, DISCARD, PURIFY, DELIVER\}, which delivers Werner pairs
    into an external link buffer. Bottom: the sequential
    (swap-and-wait) and synchronized (wait-and-swap) network-layer
    controllers.}
  \Description{A two-row schematic. The top row shows a four-link chain
    A, s1, s2, s3, B with a link controller on each link feeding
    external link buffers. The bottom row shows sequential swapping with
    partial-chain storage and synchronized swapping after all links
    supply usable pairs.}
  \label{fig:architecture}
\end{figure*}

We model the repeater chain using two control layers, illustrated in
Fig.~\ref{fig:architecture}. At the link layer, an independently trained
controller on each elementary link creates and purifies entangled pairs
and delivers pairs that meet a fixed fidelity requirement. Each link
operates at periodic decision intervals, with links running concurrently
and raw entanglement generation attempted when the controller selects
\textsc{wait}. At the network layer, a swapping discipline determines
how those delivered pairs are stored and combined into end-to-end
entanglement. This follows the layered organization used in
quantum-network architectures~\cite{dahlberg2019link} and the
link-level-pair abstraction of the packet-switched
design~\cite{bacciottini2025internet}. Sequential and synchronized use the
same trained link controllers, physical model, fidelity requirements,
buffer capacity, and pair-selection rule; they differ only in how the
network layer stores and combines delivered entanglement.

\subsection{System Model and Physics}
\label{sec:methods-model}

The chain contains $n$ elementary links connecting $n+1$ nodes along one
fixed path. A generation attempt on a link of length $L$ takes
$\tick=L/c$, the heralding time across the link, and succeeds with
probability $\pgen=K\exp(-L/\Latt)$. For a homogeneous chain, $\tick$ is
also the network-cycle duration. In the heterogeneous chains of
Section~\ref{sec:results-heterogeneous}, link $\ell$ has period
$\tick_\ell=L_\ell/c$; the simulator advances in steps of the shortest
link period, and each link acts only at multiples of its own period. For
the link lengths considered here, all link periods are integer multiples
of the global step.

The model distinguishes two stages of quantum storage. Internal memories,
with coherence time $\Tc^{\mathrm{int}}$, are used by the link controller
while it creates, purifies, and holds elementary pairs. After delivery,
pairs enter external network-layer storage with coherence time $\Text$.
Each elementary link has an external \emph{link buffer} of capacity $B$
that records the pair fidelity and delivery time,
$(F_{\mathrm{del}},t_{\mathrm{del}})$. Only pairs that satisfy the
per-link fidelity requirement of Section~\ref{sec:methods-budget} are
admitted, and $\Fdel$ denotes their mean delivered fidelity. Sequential
additionally uses $n-1$ external \emph{chain buffers} for partial chains;
synchronized stores no partial chains.

All entangled pairs are modeled as Werner states~\cite{werner1989}, so
their quality is described by a single fidelity $F$. A pair stored for
time $\Delta t$ in a memory with coherence time $T$ decays
as~\cite{nielsen2010quantum}
\begin{equation}
  F(\Delta t)
  =
  \tfrac{1}{4}
  +\left(F-\tfrac{1}{4}\right)e^{-2\Delta t/T},
  \label{eq:decoherence}
\end{equation}
with $T=\Tc^{\mathrm{int}}$ inside the link controller and $T=\Text$ in
external storage. Swapping Werner pairs with fidelities $F_1$ and $F_2$ ~\cite{zukowski1993swapping, briegel1998repeaters},
produces a Werner pair with fidelity
\begin{equation}
  F_{\mathrm{swap}}(F_1,F_2)
  =
  F_1F_2+\frac{(1-F_1)(1-F_2)}{3}.
  \label{eq:swap}
\end{equation}

Defining $x=(4F-1)/3$ makes Eq.~(\ref{eq:swap}) multiplicative:
$x_{\mathrm{swap}}=x_1x_2$. Without storage decay between swaps,
$x_{\mathrm{e2e}}=\prod_i x_i$, independent of swap order. Thus, under
the Werner-state and ideal-operation assumptions, differences between the
two disciplines arise from which delivered pairs are selected and how
long they remain in storage, rather than from the swap order itself. The
same multiplicative form gives the fidelity allocation used in
Section~\ref{sec:methods-budget}. Local operations are treated as
instantaneous and noiseless.

These assumptions isolate the interaction between probabilistic
generation, storage decoherence, and swap timing. We neglect finite gate
times, classical-message delays beyond the link-attempt period, imperfect
local operations, and hardware-specific noise, and we idealize the
heralding of purification outcomes as described in
Section~\ref{sec:methods-link}. Such effects could alter the rates and
fidelities supplied to the network layer and therefore shift the observed
operating regimes. The simulations test the analytical predictions within
this model rather than against a specific hardware platform.

Our application objective is the asymptotic six-state QKD secret-key
fraction~\cite{bruss1998sixstate, lo2001proof}. For a Werner pair of
fidelity $F$,
\begin{equation}
  \begin{aligned}
    r(F) &= \max\{0,\;1-H(F)\},\\
    H(F) &= -F\log_2F
      -(1-F)\log_2\tfrac{1-F}{3}.
  \end{aligned}
  \label{eq:keyfraction}
\end{equation}
The key fraction becomes positive above
$\Fmin\approx0.8114$. For a Werner state, the quantum bit error rate is
$Q=2(1-F)/3$, so this threshold corresponds to
$Q\approx12.6\%$. All fidelity requirements below are derived from this
threshold.

Equation~(\ref{eq:keyfraction}) depends only on fidelity because a Werner
state has the same error rate in all three six-state measurement bases.
The pooled estimate of Section~\ref{sec:methods-eval} relies on the same
symmetry. With basis-dependent noise, fidelity alone would not determine
the key fraction and a basis-resolved security analysis would be required.
We also neglect finite-length effects, which depend on block size and
observed error statistics.

\subsection{Fidelity Budget and Storage Cutoffs}
\label{sec:methods-budget}

Because swaps multiply $x$, $m$ identical pairs joined without
intervening decay reach $\Fmin$ if each pair has fidelity at least
\begin{equation}
  \Freq(m)
  =
  \tfrac{1}{4}+\tfrac{3}{4}x_{\min}^{1/m},
  \qquad
  x_{\min}=\tfrac{4\Fmin-1}{3}.
  \label{eq:freq}
\end{equation}

We use this expression to assign each elementary link an equal share of
the end-to-end fidelity budget. For $n=4$, $\Freq(4)=0.9476$. We assume
a raw-generation fidelity of $F_0=0.94$, which is slightly below
$\Freq(4)$, so the link controller must purify before delivery.

A partial chain spanning $\ell$ links receives the corresponding storage
fidelity threshold
\[
  F_{\mathrm{bar},\ell}=\Freq(n/\ell),
  \qquad
  x_{\mathrm{bar},\ell}=x_{\min}^{\ell/n}.
\]

Thus, an $\ell$-link object accounts for $\ell/n$ of the total multiplicative
fidelity budget in $x$. The equivalent argument $n/\ell$ in
$\Freq(n/\ell)$ need not be an integer. We retain the notation ``bar''
for this threshold in the algorithms. Table~\ref{tab:budget}
lists the fidelity thresholds for the $n=4$ chain, including the
emitted end-to-end pair.

\begin{table}[htbp]
  \caption{Fidelity thresholds for stored objects and the emitted
    end-to-end pair on the $n=4$ chain, from Eq.~(\ref{eq:freq}).}
  \label{tab:budget}
  \begin{tabular}{lcc}
    \toprule
    Object & Allocated requirement & Fidelity threshold\\
    \midrule
    Link pair ($\ell=1$)    & $\Freq(4)$       & $0.9476$\\
    2-link chain ($\ell=2$) & $\Freq(2)$       & $0.8989$\\
    3-link chain ($\ell=3$) & $\Freq(4/3)$     & $0.8536$\\
    Emitted pair ($\ell=4$) & $\Freq(1)=\Fmin$ & $0.8114$\\
    \bottomrule
  \end{tabular}
\end{table}

This allocation is preserved under swapping. If objects spanning $a$ and
$b$ links satisfy their respective thresholds, then
\[
  x_{\mathrm{out}}
  =x_ax_b
  \geq x_{\min}^{a/n}x_{\min}^{b/n}
  =x_{\min}^{(a+b)/n},
\]
so their output satisfies the threshold for an $(a+b)$-link object.
Consequently, extending an $\ell$-link chain with a valid elementary pair
produces an object satisfying the $(\ell+1)$-link threshold, and the final
extension reaches at least $\Fmin$. Checking each stored object against
its threshold when it is created and again before use therefore guarantees
that an emitted end-to-end pair remains above the QKD threshold.

This allocation is sufficient but not necessary. An object below its
assigned threshold could still contribute to a valid end-to-end pair if
another part of the chain had excess fidelity. We do not redistribute
such excess; objects below their assigned threshold are discarded.

Equation~(\ref{eq:decoherence}) converts each fidelity threshold into a
storage lifetime. An object created at fidelity $F$ reaches
$F_{\mathrm{bar}}$ after
\begin{equation}
  \tcut
  =
  \frac{\Text}{2}
  \ln\!\frac{F-\tfrac{1}{4}}
           {F_{\mathrm{bar}}-\tfrac{1}{4}},
  \label{eq:tcut}
\end{equation}
which we call its \emph{storage cutoff}. Beyond $\tcut$ \cite{inesta2023optimal}, the object falls
below its allocated fidelity threshold and is no longer used. If its
fidelity is already at or below the threshold when created, its cutoff is
nonpositive and the object is rejected.

For an elementary pair, the cutoff clock begins at delivery. For a
partial chain, it begins at the swap that formed the chain, because decay
before that swap is already included in its merged fidelity. A link pair
moved into the first chain buffer keeps its original delivery time.
Expired objects are removed before the network controller acts, and
candidates are checked again immediately before use. Both disciplines use
the same fidelity thresholds and cutoff rule.

\subsection{Link Layer: Reinforcement-Learned Delivery}
\label{sec:methods-link}

Each elementary link runs an independent reinforcement-learning controller
based on the WN2M2 environment of Yau et al.~\cite{yau2025rl}. The
environment contains two memories at each endpoint of the link and four
actions: \textsc{wait}, \textsc{discard}, \textsc{purify}, and
\textsc{deliver}. A successful generation attempt produces a Werner pair,
and \textsc{purify} applies one round of DEJMPS
distillation~\cite{dejmps1996}. We use the WN2M2 link dynamics while
training our own policies for the fidelity requirements and QKD objective
used here.

The controller observes $s=(F_1,F_2,p,t)$, where $F_1$ and $F_2$ are the
fidelities in its two memories, $p\in(0,1]$ is the success probability
associated with the most recent unheralded purification, and
$t=T_{\mathrm{elapsed}}/(10^5\tick)$ is the normalized episode time.
The factor $10^5\tick$ sets a fixed time scale for the controller's
input, and the same normalization is used during training and evaluation.
An episode consists of the sequence of controller actions from reset
until a successful delivery or truncation after $1000$ actions.
The \textsc{wait} action attempts entanglement generation,
\textsc{discard} removes a stored pair, \textsc{purify} combines the
two stored pairs, and \textsc{deliver} passes the selected pair to the
network layer. A successful delivery starts a new episode.
In the link-layer model, each \textsc{wait} action advances the
physical clock by one generation-attempt period $\tick$,
while \textsc{discard}, \textsc{purify}, and \textsc{deliver}
are treated as instantaneous.

A purified pair may be delivered before its purification outcome is
heralded. If its current success probability is $p<1$, the simulator
samples that outcome at delivery: with probability $p$ the pair enters
the network-layer link buffer, and otherwise the delivery fails and
the episode continues. Only successful pairs therefore reach the network
layer. This idealization omits the cost of a failed pair that, in a real
network, could already have been used before its purification failure was
known; the same convention is applied to both disciplines.

Actions that are physically unavailable in the current state are masked
before the softmax. The policy is a neural network with two hidden layers
($4\!\to\!64\!\to\!64\!\to\!4$) followed by a masked softmax output.
This replaces the Fourier-linear policy of the original framework while
retaining the same decision process.

Training uses REINFORCE~\cite{williams1992simple, sutton2018rl} with
Adam~\cite{kingma2015adam}, a learning rate of $2.5\times10^{-4}$,
gradient-norm clipping at $1$, and baselines given by the batch mean
fidelity return and the per-step mean time-to-go. Each training iteration
contains a batch of $10^4$ episodes, after which one policy update is
performed. All controllers begin from the same randomly initialized
network, with the random-number generators seeded.

For each training iteration, let $J_F$ denote the mean fidelity return
over its batch of $10^4$ episodes and let $J_T$ denote the corresponding
mean elapsed episode time in seconds. Failed or below-requirement
deliveries, as well as episodes ending without a successful delivery, are
assigned zero fidelity return. Thus, $J_F$ is a batch-level training
quantity and is not the mean fidelity conditioned on successful delivery.
Once the batch reaches the key-producing region, the update is based on
the QKD utility
\[
  u=\frac{\max\{0,1-H(J_F)\}}{J_T}.
\]

Direct optimization of this objective fails when
$\tfrac14<J_F<\Fmin$. In this region $u=0$, so the objective provides
neither a signal favoring higher fidelity nor a penalty for longer
episodes. Rewarding fidelity directly supplies the first signal, but
without the second the policy can fall into an \emph{over-purification
trap}: when decoherence is slow, the controller may continue purifying
rather than reaching a useful delivery policy.

To provide both signals in this region, apart from the numerical edge
cases below, the policy update combines a fidelity-gradient contribution
weighted by $1/\max(J_T,10^{-6})$ with a time-gradient contribution
weighted by
\begin{equation}
  \left.\frac{\partial u}{\partial J_T}\right|_{\mathrm{surrogate}}
  =
  -c\,\frac{1-H(\Freq(n))}{J_T^2},
  \qquad c=0.5.
  \label{eq:lenpen}
\end{equation}
The reference $\Freq(n)$ depends only on the chain length and the QKD
threshold. These weights are a training device rather than derivatives of
the flat QKD objective in this region.

The update rules are applied in order. If $J_T<10^{-10}$,
$J_F<0.250001$, or $J_F>0.999999$, only the fidelity term is used, with
weight $1/\max(J_T,10^{-6})$. Otherwise, if $J_F<\Fmin$, the same
fidelity weight is combined with the surrogate time term in
Eq.~(\ref{eq:lenpen}). Once $J_F\geq\Fmin$, the true derivatives of the
QKD objective are used.

We chose $c=0.5$ after comparing three values. With $c=0$, the
over-purification trap remains and training settles near
$J_F\approx0.887$. With $c=1$, the time pressure is too strong and
training stalls near $J_F\approx0.51$ without delivering above the
required fidelity. With $c=0.5$, training escapes the trap at every
coherence setting. This is an empirical choice rather than a systematic
sensitivity study. Although the surrogate penalty is inactive once the
key fraction becomes positive, it can still affect which policy training
reaches.

Training stops when the most recent $70$ iterations all have positive
key with $J_F\geq\Freq(n)$, the range of their $J_F$ values is below
$10^{-3}$, and the absolute least-squares slope multiplied by the window
length is below $3\times10^{-4}$, or when a hard iteration cap is reached.
All but one of the $24$ configurations satisfy this criterion. The
exception, at $L=10$~km and $\Tc^{\mathrm{int}}/\tick=250$, reaches the
$1000$-iteration cap and has the lowest mean delivered fidelity in the
policy bank. We retain it in the network evaluation and identify it as
capped. The stopping rule tests whether training has settled; it does not
establish global optimality.

We train one policy for each link length and internal-coherence setting
used in the evaluation. During evaluation, the controller takes its most
probable action rather than sampling from the policy. The same trained
policy is used for the corresponding link under both network-layer
disciplines.

\subsection{Network-Layer Swapping Disciplines}
\label{sec:methods-controllers}

Both disciplines use the same cycle order. First, each link whose next
decision is due takes one controller action, and any successfully
delivered pair enters its link buffer. Next, expired objects are removed
and link-buffer occupancy is recorded. Finally, the network-layer
controller acts. Because new deliveries enter the buffers before this
last step, a pair delivered during the current cycle can be used
immediately at zero external-storage age. Synchronized can therefore
produce an end-to-end pair even when the storage cutoff is shorter than
one cycle, provided all links deliver usable pairs in that same cycle.

In the network simulation, each link takes one action per scheduled link
step, regardless of the action. This differs from the link-training clock,
where only \textsc{wait} advances physical time. Internal pairs therefore
decay only on \textsc{wait}; non-\textsc{wait} actions occupy a network
decision step without additional internal-memory decay. This simulator
convention is applied identically under both disciplines.

Whenever a buffer contains several candidates, the controller chooses the
freshest one. For a link buffer this is the most recently delivered pair;
for a chain buffer it is the partial chain whose oldest constituent pair
is most recent. Once a partial chain has been formed, later elementary-link
deliveries do not replace any of its constituents; they may instead be
used to form or extend other chains. When a buffer is full, its oldest
entry is dropped. Because storage cutoffs depend on fidelity as well as
age, the oldest entry is not necessarily the one closest to expiry.
Selected entries are checked again for expiry and decayed to the current
time before use.

\paragraph{Sequential swapping (swap-and-wait).}
Sequential keeps one chain buffer $C_\ell$ for each partial-chain length
$\ell=1,\ldots,n-1$. Each $C_\ell$ may contain multiple partial chains
spanning $\ell$ links, up to the buffer capacity $B$. During a cycle the
controller extends stored chains whenever a pair from the next elementary
link is available, and a chain may advance through several links in the
same cycle, since entanglement generation proceeds concurrently across
the elementary links. Once a chain spans all $n$ links, it is emitted as
an end-to-end pair.

Newly seeded chains enter $C_1$ only after the extension pass. A pair
used to start a new chain therefore cannot be extended until the following
cycle, making one-cycle survival necessary for this implementation.
Sequential may emit more than one end-to-end pair in a cycle when its
buffers are sufficiently populated.

\begin{algorithm}[htbp]
  \caption{Sequential (swap-and-wait) controller, one cycle at time
    $t$. $D$ is Eq.~(\ref{eq:decoherence}) at $\Text$ and
    $F_{\mathrm{swap}}$ is Eq.~(\ref{eq:swap}).}
  \label{alg:seq}
  \begin{algorithmic}[1]
    \Require link buffers $B_1,\ldots,B_n$;
             chain buffers $C_1,\ldots,C_{n-1}$
    \State $C_\ell.\textsc{DiscardExpired}(t)$ for all $\ell$
           \Comment{age from $t_{\mathrm{sw}}$}
    \For{$\ell=2,\ldots,n$}
      \While{$C_{\ell-1}$ and $B_\ell$ both non-empty}
        \State $(F_{\mathrm{ch}},t_{\mathrm{sw}},t_{\mathrm{old}})
          \gets C_{\ell-1}.\textsc{PopFreshest}()$
        \If{$t-t_{\mathrm{sw}}>\tcut$ of the entry}
          \State \textbf{continue}
        \EndIf
        \State $(F_\ell,t_\ell)
          \gets B_\ell.\textsc{PopFreshest}()$
        \If{$t-t_\ell>\tcut$ of the entry}
          \State restore the chain to $C_{\ell-1}$ without
                 resetting its age
          \State \textbf{continue}
        \EndIf
        \State $F'_{\mathrm{ch}}
          \gets D(F_{\mathrm{ch}},t-t_{\mathrm{sw}})$
        \State $F'_\ell\gets D(F_\ell,t-t_\ell)$
        \State $F_{\mathrm{new}}
          \gets F_{\mathrm{swap}}(F'_{\mathrm{ch}},F'_\ell)$
        \State $t'_{\mathrm{old}}
          \gets\min(t_{\mathrm{old}},t_\ell)$
        \If{$\ell=n$}
          \State \textbf{emit} $F_{\mathrm{new}}$ at time $t$
        \Else
          \State $C_\ell.\textsc{Push}
            (F_{\mathrm{new}},t_{\mathrm{sw}}=t,t'_{\mathrm{old}})$
            \Comment{threshold $\Freq(n/\ell)$}
        \EndIf
      \EndWhile
    \EndFor
    \While{$B_1$ non-empty} \Comment{seed after extension}
      \State $(F_1,t_1)\gets B_1.\textsc{PopFreshest}()$
      \If{$t-t_1\leq\tcut$ of the entry}
        \State $C_1.\textsc{Push}(F_1,t_{\mathrm{sw}}=t_1,t_1)$
      \EndIf
    \EndWhile
  \end{algorithmic}
\end{algorithm}

\paragraph{Synchronized swapping (wait-and-swap).}
Synchronized retains no partial chains. Each link buffer may contain
multiple delivered elementary pairs. When every link has a usable pair,
the controller takes one pair from each buffer and joins that set into one
end-to-end pair using a balanced binary tree of swaps. Under the
instantaneous-operation model, the $\lceil\log_2 n\rceil$ swap rounds
complete within the same cycle. Because Werner-state swapping is
multiplicative in $x$, the tree structure does not change the final
fidelity in this model.

The controller performs at most one such synchronized swap per cycle,
even if another complete set remains in the link buffers; this is a
restriction of the discipline evaluated here. Every selected elementary
pair must still satisfy $\Freq(n)$ when used, so joining the $n$ pairs
without further storage yields an end-to-end pair meeting $\Fmin$.

\begin{algorithm}[htbp]
  \caption{Synchronized (wait-and-swap) controller, one cycle at
    time $t$. \textsc{PopFreshestLive} discards expired candidates
    until it finds a live entry or returns $\varnothing$.
    $D$ is Eq.~(\ref{eq:decoherence}) at $\Text$.}
  \label{alg:sync}
  \begin{algorithmic}[1]
    \Require link buffers $B_1,\ldots,B_n$
    \If{any $B_\ell$ is empty}
      \State \Return
    \EndIf
    \For{$\ell=1,\ldots,n$}
      \State $q_\ell\gets B_\ell.\textsc{PopFreshestLive}(t)$
      \If{$q_\ell=\varnothing$}
        \State \Return \Comment{earlier pops are not restored}
      \EndIf
      \State $(F_\ell,t_\ell)\gets q_\ell$
      \State $\tilde{F}_\ell\gets D(F_\ell,t-t_\ell)$
    \EndFor
    \State \textbf{emit}
      $\textsc{SyncSwap}([\tilde{F}_1,\ldots,\tilde{F}_n])$
      at time $t$
    \Statex
    \Function{SyncSwap}{$[F_1,\ldots,F_m]$}
      \If{$m=1$}
        \State \Return $F_1$
      \EndIf
      \State $\mathit{mid}\gets\lfloor m/2\rfloor$
      \State $F_{\mathrm{left}}\gets
        \textsc{SyncSwap}([F_1,\ldots,F_{\mathit{mid}}])$
      \State $F_{\mathrm{right}}\gets
        \textsc{SyncSwap}([F_{\mathit{mid}+1},\ldots,F_m])$
      \State \Return
        $F_{\mathrm{swap}}(F_{\mathrm{left}},F_{\mathrm{right}})$
    \EndFunction
  \end{algorithmic}
\end{algorithm}

Synchronized is distinct from SWAP-ASAP, where each repeater swaps
locally as soon as both adjacent links are
ready~\cite{inesta2023optimal, coopmans2022improved}. Synchronized uses a
global trigger and waits for the entire chain, whereas sequential
preserves partial chains between cycles.

Algorithms~\ref{alg:seq} and~\ref{alg:sync} describe one network cycle.
When $C_\ell.\textsc{Push}$ is called, the new chain receives the
corresponding cutoff from Section~\ref{sec:methods-budget} and is rejected
if that cutoff is nonpositive. Candidates are checked again before use
because their remaining lifetimes can differ. Neither algorithm separately
checks the final end-to-end fidelity; the threshold allocation and expiry
checks of Section~\ref{sec:methods-budget} guarantee that an emitted pair
remains above $\Fmin$. Diagnostic bookkeeping is omitted from the
pseudocode.

\paragraph{Shared settings and finite buffers.}
Both disciplines use the same fidelity thresholds, storage-cutoff rule,
freshest-first selection, cycle order, and buffer capacity $B$.
Sequential additionally has chain buffers, so equal capacity per buffer
does not imply equal total memory use.

We set $B=20$ independently for every link buffer and, under sequential,
every chain buffer. Mean link-buffer occupancy remains below this value:
the largest observed mean is about $11$ in the heterogeneous study, while
homogeneous means remain below $3.2$. These averages do not establish
that buffers never saturate, because we did not record full occupancy
distributions or overflow-drop counts, nor do they characterize peak
occupancy in the chain buffers. Our results therefore describe operation
at $B=20$, rather than a verified overflow-free memory size. We do not
vary the pair-selection rule or cycle order.

\subsection{Evaluation Protocol and Metric}
\label{sec:methods-eval}

For a trial that delivers at least one end-to-end pair, we define the
pooled six-state secret-key-rate utility
\begin{equation}
  \uskr
  =
  \frac{\max\{0,\;1-H(\bar{F}_{\mathrm{e2e}})\}}
       {\langle\Delta t\rangle},
  \qquad
  \langle\Delta t\rangle=\frac{T_{\mathrm{last}}}{N},
  \label{eq:uskr}
\end{equation}
where $\bar{F}_{\mathrm{e2e}}$ is the mean fidelity of the delivered
end-to-end pairs, $N$ is the number delivered, and $T_{\mathrm{last}}$
is the time from the start of the trial to the last delivery. A trial
with no deliveries has zero utility. The interval includes the initial
wait before the first delivery but excludes time after the last one.
Thus, $1/\langle\Delta t\rangle=N/T_{\mathrm{last}}$ measures the
delivery rate of the observed stream rather than throughput over the
full simulation window, which would use $N/T_{\mathrm{sim}}$. Both
disciplines use the same definition.

We apply the key fraction to the pooled fidelity rather than averaging
per-pair key fractions. Because $H$ is concave,
$H''(F)=-(\ln2)^{-1}[F^{-1}+(1-F)^{-1}]<0$, Jensen's inequality implies
that averaging per-pair key fractions would give a larger value than
applying the key fraction to the mean fidelity. Our metric instead treats
the delivered pairs as one combined key block. For Werner states, the
QBER is linear in $F$, so the pooled fidelity corresponds exactly to the
average error rate of that block. We therefore pool the delivered pairs
rather than sorting them by fidelity and assigning separate key fractions.

Across trials that deliver, fidelities are weighted by the number of
deliveries, and the corresponding mean delivery interval is
$\sum_jT_{\mathrm{last},j}/\sum_jN_j$. We report $95\%$ confidence
intervals from the per-trial utilities. Because the key fraction is
nonlinear, the rate computed from pooled quantities and the mean of the
per-trial utilities need not coincide exactly. At every evaluated
setting, either all $200$ trials deliver at least one end-to-end pair or
none do.

The link-layer secret-key rate reported in
Section~\ref{sec:results-training} uses the same form as
Eq.~(\ref{eq:uskr}) but follows the link-training clock, where only
\textsc{wait} contributes elapsed time. It therefore characterizes an
isolated link and is not directly comparable in magnitude with the
network-level rate. When link-delivery statistics are used to predict
network-level behavior, episode lengths are instead counted in
network-simulation cycles.

We also record mean link-buffer occupancy once per cycle, after expired
objects are removed and before the network controller acts, averaged over
cycles and elementary links. It counts entries present at that sampling
point, some of which may later be discarded by a validity check. Chain
buffers are excluded, so this quantity is not a measure of total
quantum-memory use.

We use an attenuation length $\Latt=22$~km, propagation speed
$c=2\times10^5$~km/s, and coupling and detection factor $K=0.9$. The
homogeneous study considers $L=5$ and $10$~km, corresponding to attempt
periods of $25$ and $50~\mu$s and generation probabilities of
approximately $0.72$ and $0.57$, respectively. For each link length, the
controllers are trained at
\[
  \Tc^{\mathrm{int}}/\tick
  \in
  \{5,10,25,50,100,250,500,1000,2500,5000,10^4,2\times10^4\}.
\]
During network evaluation, $\Tc^{\mathrm{int}}=\Text$ at each operating
point.

Each network configuration is evaluated using $200$ independent trials
of duration $T_{\mathrm{sim}}=5$~s per discipline. For a given operating
point, trial $i$ uses the same random seed under sequential and
synchronized. Because the link controllers do not observe the
network-layer discipline, the two simulations receive the same
realization of link deliveries for that trial and differ only in how
those deliveries are stored and swapped. Confidence intervals are
reported separately for the two disciplines and do not exploit this
pairing. The simulator is implemented in Python, using NumPy for the
physical model, PyTorch for the policies, and Gymnasium for the link
environment.
\section{Results}
\label{sec:results}

We first characterize the trained link controllers in isolation and then
compare sequential and synchronized swapping on the four-hop repeater
chain. Network-level performance is reported using the pooled six-state
secret-key-rate utility $\uskr$ of Eq.~(\ref{eq:uskr}) and the evaluation
protocol of Section~\ref{sec:methods-eval}. Where shown, error bars denote
$95\%$ confidence intervals and are often smaller than the plot markers.

Section~\ref{sec:results-training} examines the link-layer policies across
the coherence range. Section~\ref{sec:results-homogeneous} compares the
two swapping disciplines on homogeneous chains with $5$ and $10$~km
links, and Section~\ref{sec:results-heterogeneous} introduces one slower
$10$~km link into an otherwise $5$~km chain and varies its position.

\subsection{Link-Layer Training}
\label{sec:results-training}

\begin{figure}[htbp]
  \centering
  \includegraphics[width=\linewidth]{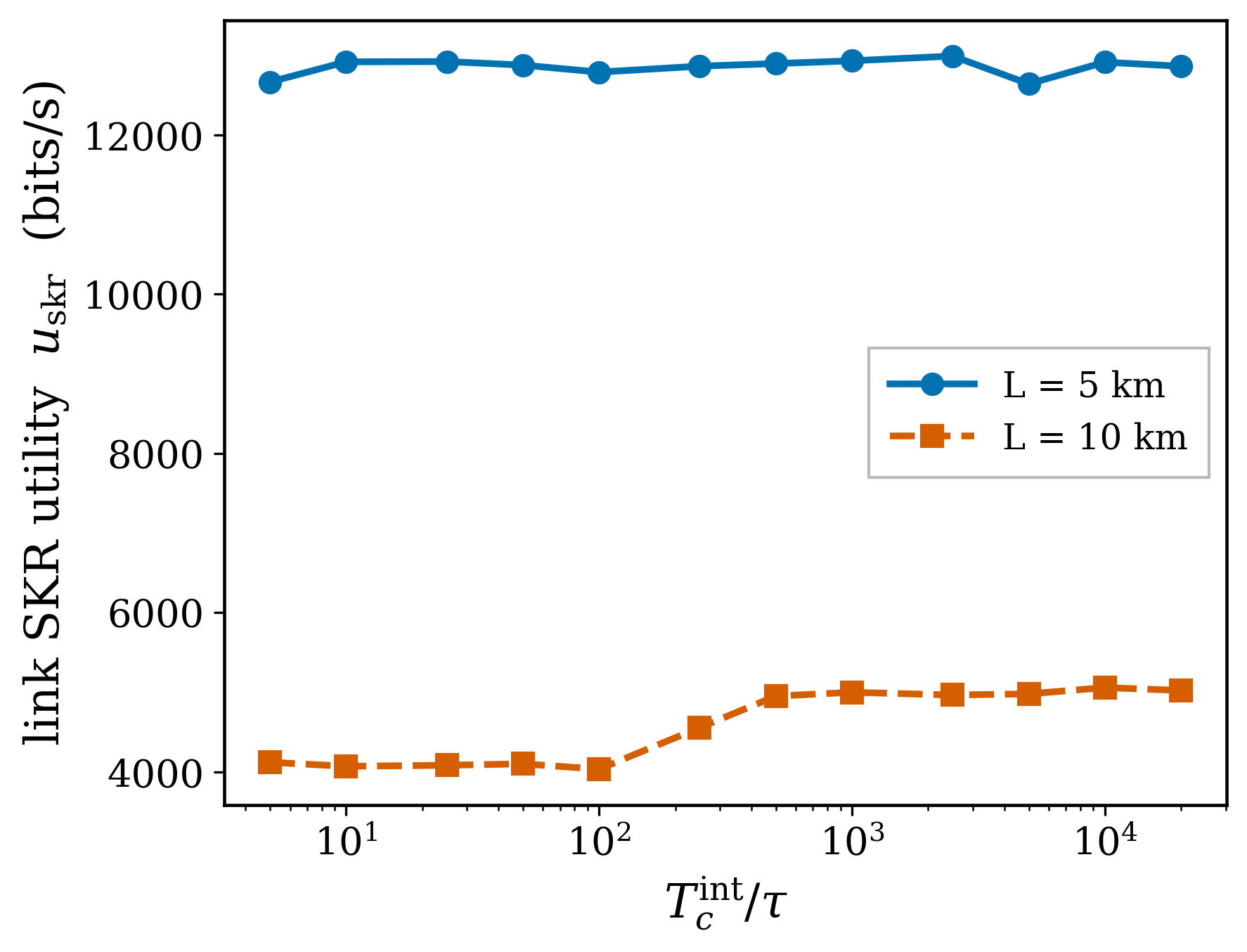}
  \caption{Link-layer secret-key-rate utility $\uskr$ (bits/s) of the
  trained WN2M2 link controllers versus the coherence-to-cycle ratio
  $\Tc^{\mathrm{int}}/\tick$, for $L=5$ and $L=10$~km. Each marker
  corresponds to one controller trained and evaluated at that setting.}
  \Description{Two curves of link secret-key rate against the internal
    coherence-to-cycle ratio on a log x-axis; the shorter link sustains a
    higher and flatter rate than the longer link.}
  \label{fig:training-uskr}
\end{figure}

\begin{figure}[htbp]
  \centering
  \includegraphics[width=\linewidth]{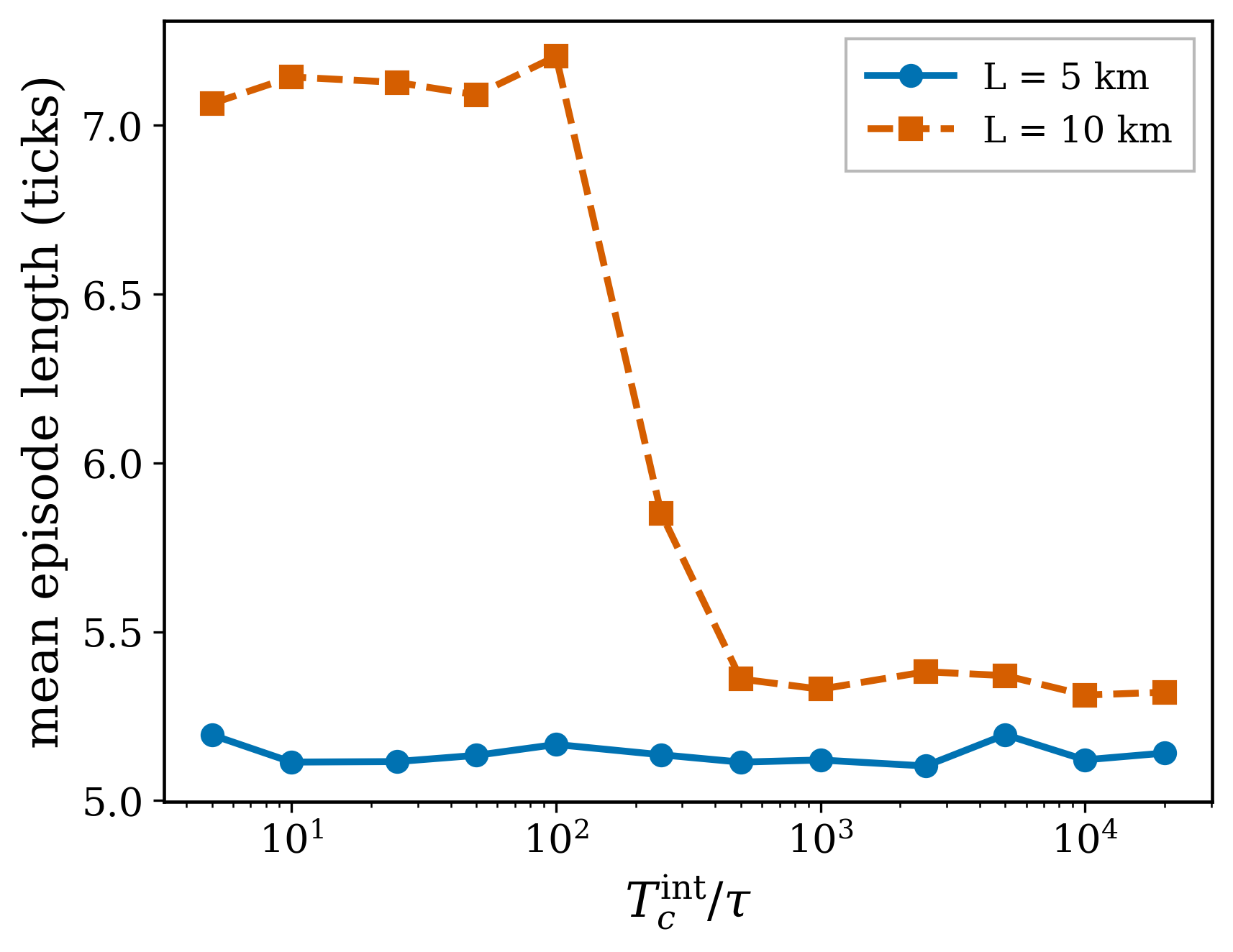}
  \caption{Mean episode length, expressed in network-simulation cycles,
  of the trained link controllers versus $\Tc^{\mathrm{int}}/\tick$,
  for $L=5$ and $L=10$~km. Same controllers as
  Fig.~\ref{fig:training-uskr}.}
  \Description{Two curves of mean episode length against the internal
    coherence-to-cycle ratio; the longer link drops from about seven to
    about five cycles as coherence increases, while the shorter link
    stays near five.}
  \label{fig:training-length}
\end{figure}

We first characterize the trained link controllers before placing them in
the repeater chain. Figures~\ref{fig:training-uskr}
and~\ref{fig:training-length} show their link-layer secret-key rate and
mean episode length across the coherence settings defined in
Section~\ref{sec:methods-eval}.

Figure~\ref{fig:training-uskr} shows the link-layer rate measured on the
training clock of Section~\ref{sec:methods-eval}. At $L=5$~km, the rate
remains near $12.8$~kbit/s across the full coherence range. At
$L=10$~km, it remains near $4.0$~kbit/s at short coherence, rises to
about $5.0$~kbit/s between $\Tc^{\mathrm{int}}/\tick\approx100$ and
$500$, and then levels off.

Figure~\ref{fig:training-length} shows the mean episode length in
network-simulation cycles. At $L=5$~km, it remains near $5.1$ cycles.
At $L=10$~km, it decreases from about $7.1$ to $5.3$ cycles between
$\Tc^{\mathrm{int}}/\tick\approx100$ and $500$. On the training clock,
where only \textsc{wait} advances time, the corresponding episode
durations are about $2.1$ attempt periods at $L=5$~km and
$2.7$--$3.4$ at $L=10$~km. The increase in link-layer rate at
$L=10$~km coincides with the reduction in episode length.

All trained controllers deliver pairs with mean fidelity above the
four-hop requirement $\Freq(4)=0.9476$, using about $1.08$ purification
rounds per successful delivery. Among deliveries meeting the requirement,
mean fidelities range from $0.9567$ to $0.9575$, with $20$ of the $24$
controllers reaching $0.9575$. At least $99.97\%$ of deliveries meet
the requirement for every controller except the capped run at $L=10$~km
and $\Tc^{\mathrm{int}}/\tick=250$, identified in
Section~\ref{sec:methods-link}. For this controller, $99.69\%$ of
deliveries meet the requirement, and the mean accepted fidelity is
$0.9567$. It lies within the coherence range where the $L=10$~km
episode length changes and is used at the first setting where sequential
produces key in Section~\ref{sec:results-homogeneous}.

Overall, the trained link controllers maintain nearly constant
delivered-pair fidelity across the coherence sweep, while the main
coherence-dependent change at $L=10$~km is in delivery-episode length.

\subsection{Homogeneous Chain}
\label{sec:results-homogeneous}

We next compare the two swapping disciplines on four-hop chains of
identical links over the coherence range defined in
Section~\ref{sec:methods-eval}. At each setting,
$\Tc^{\mathrm{int}}=\Text$, and both disciplines use the same trained
link controllers.

\begin{figure*}[htbp]
  \centering
  \includegraphics[height=0.44\textheight,keepaspectratio]{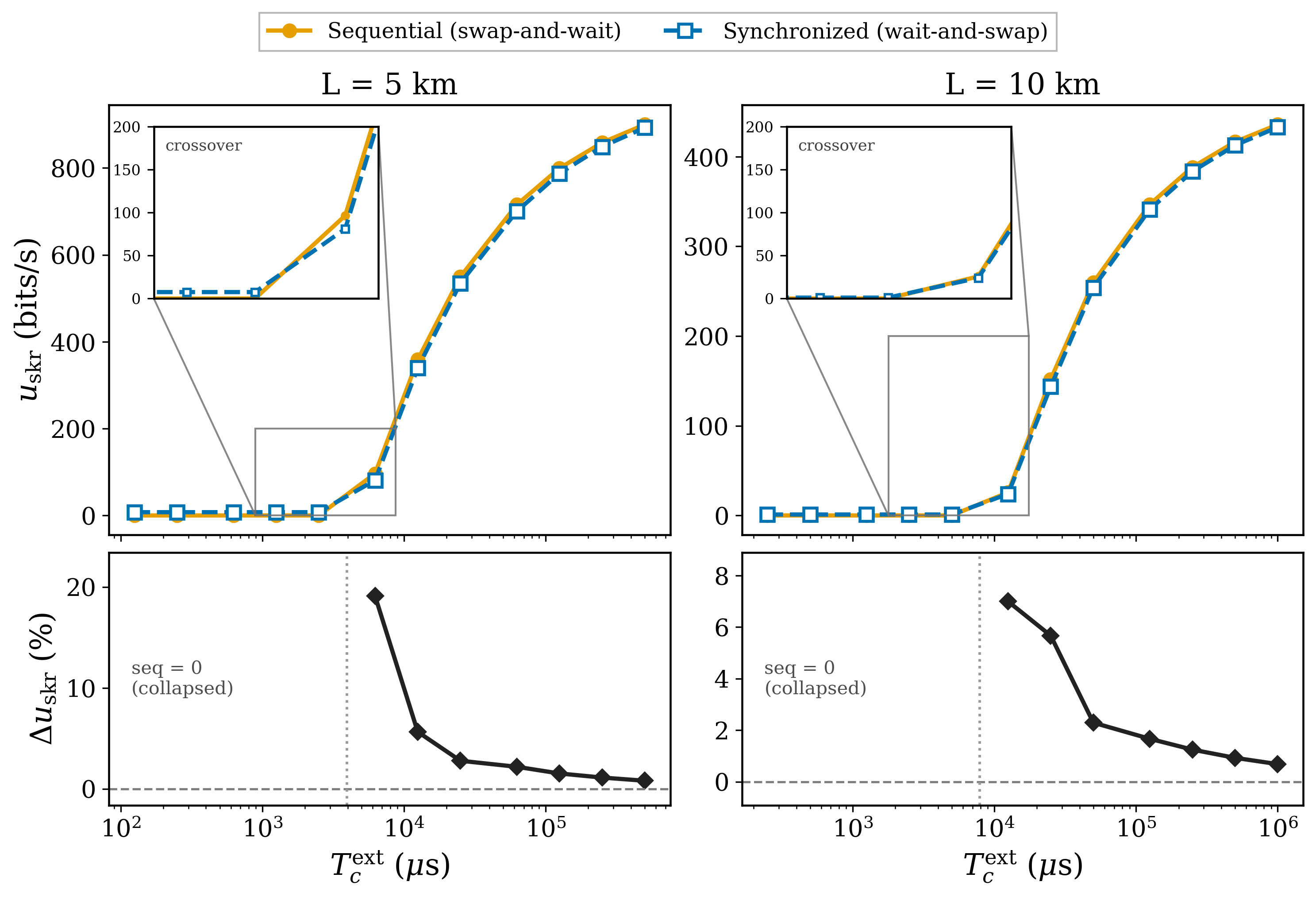}
  \caption{Six-state secret-key-rate utility $\uskr$ (bits/s) of the
    sequential (swap-and-wait) and synchronized (wait-and-swap)
    disciplines versus external storage coherence $\Text$; $L=5$~km
    (left), $L=10$~km (right). Lower panels show the relative difference
    $\Delta\uskr=(\uskr^{\mathrm{seq}}-\uskr^{\mathrm{sync}})
    /\uskr^{\mathrm{sync}}$, in percent, where both disciplines produce
    key. Settings where sequential produces no key are indicated
    separately. The vertical dotted line marks the geometric midpoint of
    the observed transition interval, not a measured threshold. Insets
    magnify the short-coherence region.}
  \Description{Secret-key-rate curves versus coherence time for two link
    lengths, with lower panels showing the sequential advantage where both
    disciplines produce key. The advantage is largest at the first such
    point and decreases as coherence increases.}
  \label{fig:crossover}
\end{figure*}

\begin{figure*}[htbp]
  \centering
  \includegraphics[height=0.31\textheight,keepaspectratio]{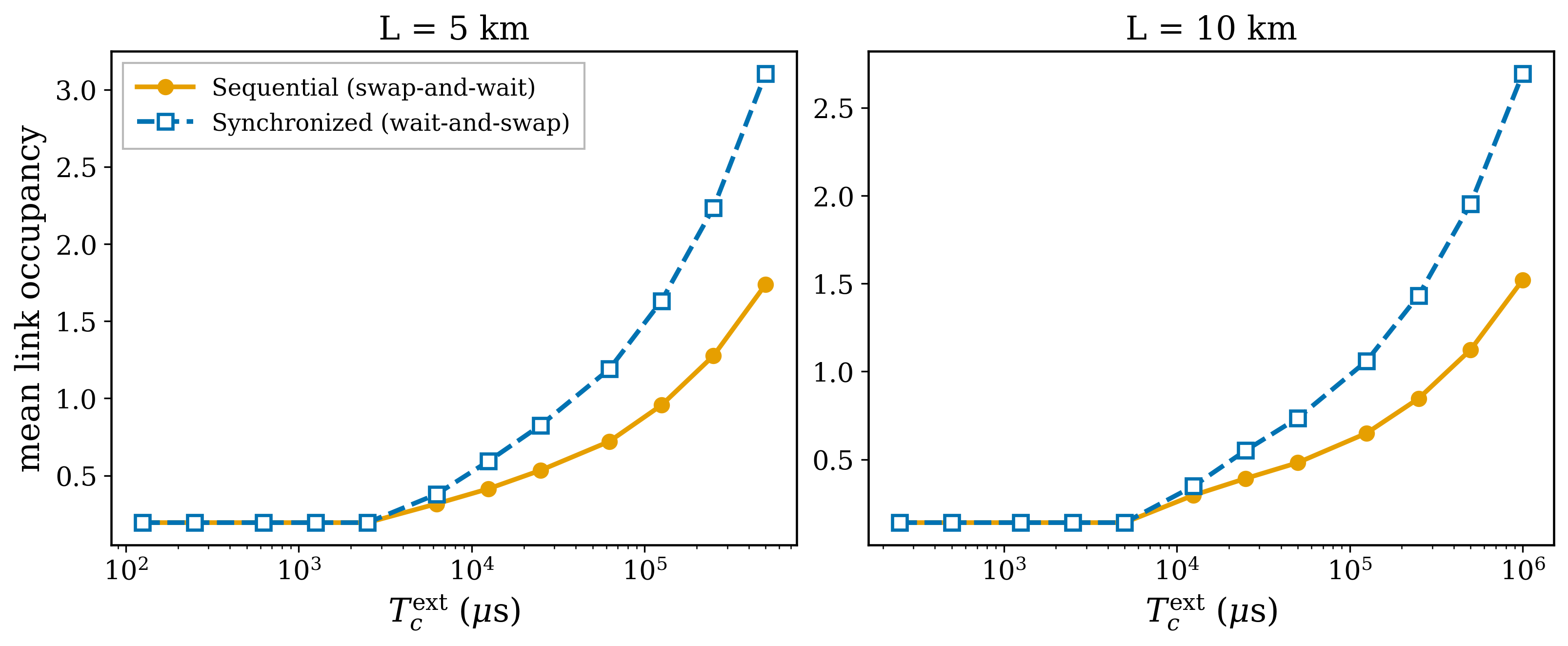}
  \caption{Mean link-buffer occupancy of the two disciplines versus
    external storage coherence $\Text$; $L=5$~km (left), $L=10$~km
    (right). Occupancy is sampled once per cycle and averaged over cycles
    and links. Link buffers have capacity $B=20$; chain buffers are not
    included.}
  \Description{Two panels of mean link-buffer occupancy against coherence
    time. Both curves are flat and equal at low coherence, then rise with
    synchronized above sequential.}
  \label{fig:occupancy}
\end{figure*}

Figure~\ref{fig:crossover} shows the secret-key rates of both disciplines.
As coherence increases, the rates rise from zero or small values to about
$900$~bits/s at $L=5$~km and $435$~bits/s at $L=10$~km. The insets
magnify the short-coherence region, where the differences are less visible
on the full scale.

At the five shortest coherence settings, none of the $200$ sequential
trials delivers an end-to-end pair. Synchronized still produces key, at
$7.49$~bits/s for $L=5$~km and $1.03$~bits/s for $L=10$~km. Sequential
first produces key between $\Tc/\tick=100$ and $250$ at both link
lengths. The dotted lines mark the geometric midpoints of these sampled
intervals, near $\Text=4000~\mu$s for $L=5$~km and $7900~\mu$s for
$L=10$~km. They are visual guides, not measured thresholds.

Once both disciplines produce key, sequential has the higher rate
throughout the homogeneous-chain sweep, with its largest advantage at
the first such setting. At $L=5$~km, sequential reaches
$96.50\pm0.17$~bits/s compared with $81.03\pm0.18$~bits/s for
synchronized, a $19.1\%$ increase; at $L=10$~km, the increase is
$7.0\%$. The uncertainties are $95\%$ confidence-interval half-widths.
The advantage decreases at every subsequent setting, reaching $0.80\%$
at $L=5$~km and $0.68\%$ at $L=10$~km at the longest coherence time.
The corresponding $L=5$~km rates are $899.86\pm0.29$ and
$892.73\pm0.28$~bits/s. The results identify three regimes: short
coherence where only synchronized produces key, a transition where
sequential gains its largest advantage, and long coherence where the
rates nearly coincide.

End-to-end fidelity remains above the six-state threshold
$\Fmin=0.8114$ whenever either discipline produces key. In the
short-memory region, synchronized delivers $\Fetoe=0.844$ at both link
lengths. At the first setting where both disciplines produce key,
sequential has lower fidelity: $0.829$ versus $0.835$ at $L=5$~km,
and $0.826$ versus $0.831$ at $L=10$~km. Both approach $0.842$ at
the longest coherence time. Thus, sequential's initial SKR advantage
does not arise from higher end-to-end fidelity.

Figure~\ref{fig:occupancy} shows mean link-buffer occupancy. At short
coherence, both disciplines have the same occupancy: $0.195$ at
$L=5$~km and $0.141$ at $L=10$~km. Occupancy rises with coherence
under both disciplines but remains higher for synchronized. At the
longest coherence time, synchronized and sequential reach mean
occupancies of $3.1$ and $1.74$, respectively, at $L=5$~km, and
$2.7$ and $1.5$ at $L=10$~km. These values exclude sequential's
partial-chain storage. At $L=5$~km and the longest coherence time,
each delivered end-to-end pair spends about $364~\mu$s in total as
a partial chain.

\subsection{Heterogeneous Chain (Bottleneck)}
\label{sec:results-heterogeneous}

We next introduce one slower link by placing a $10$~km link in an
otherwise $5$~km chain and moving it across the four hops:
$(10,5,5,5)$, $(5,10,5,5)$, $(5,5,10,5)$, and $(5,5,5,10)$.
We label these configurations hop~$1$ through hop~$4$ according
to the slow-link position. Each link uses the controller trained
for its own length. To maintain a common physical coherence time
across the mixed chain, we evaluate the eight values of $\Text$
shared by the $5$ and $10$~km controller grids, from $250~\mu$s
to $5\times10^{5}~\mu$s.

\begin{figure*}[htbp]
  \centering
  \includegraphics[height=0.30\textheight,keepaspectratio]{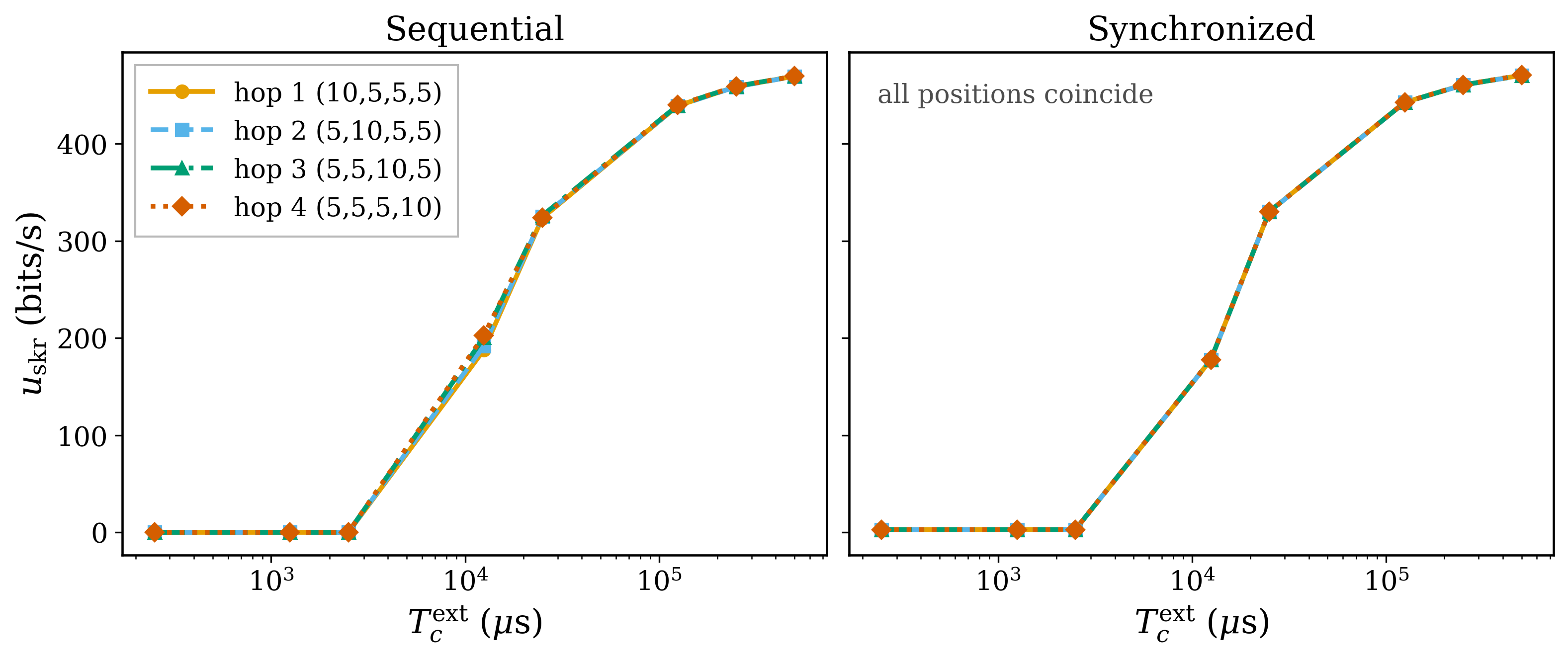}\\[2pt]
  \includegraphics[height=0.30\textheight,keepaspectratio]{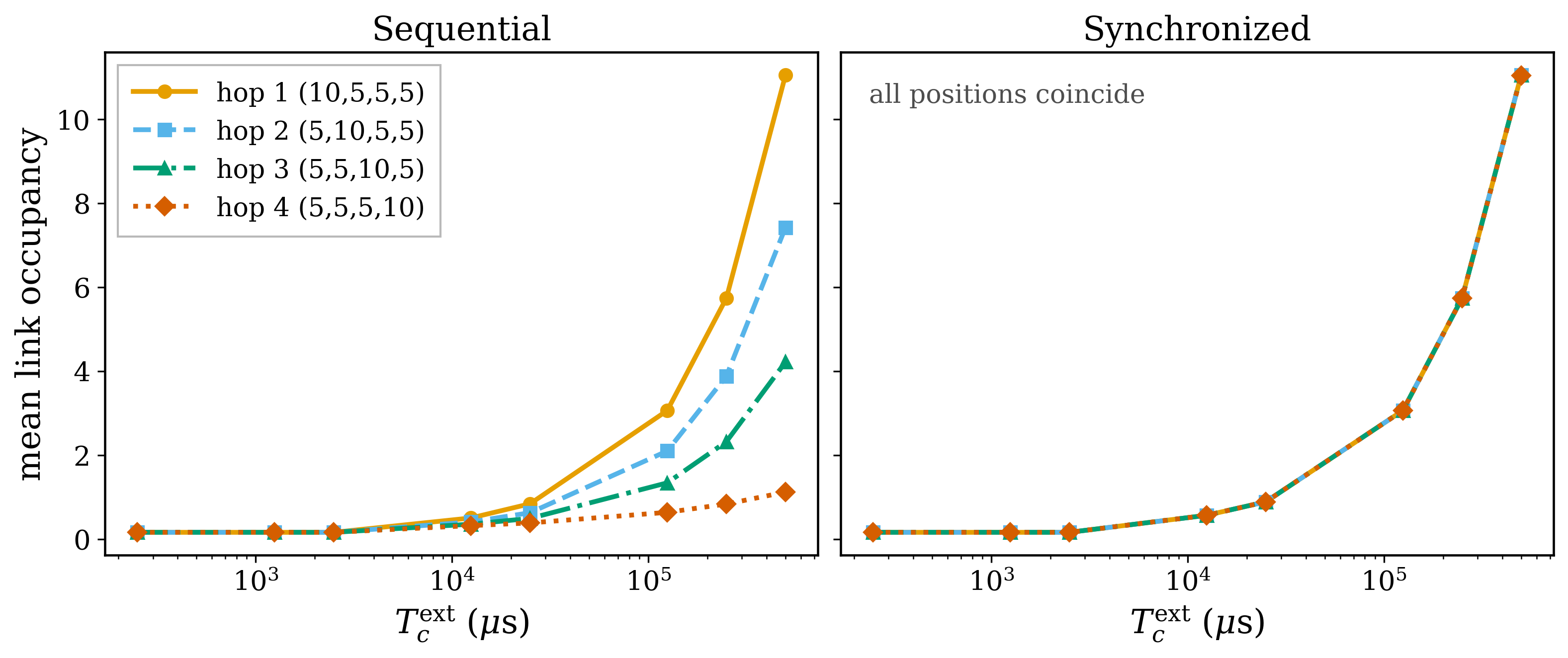}
  \caption{Heterogeneous chain with one $10$~km link in an otherwise
  $5$~km chain, for the four slow-link positions (hop~$1$ to hop~$4$),
  versus external storage coherence $\Text$. Top: secret-key-rate utility
  $\uskr$ (bits/s). Bottom: mean link-buffer occupancy, averaged over
  links and sampled once per cycle. Sequential (left), synchronized
  (right).}
  \Description{Four panels. Top: secret-key rate against coherence time;
    the four slow-link position curves overlap across most of the range,
    with a small spread among the sequential curves near the first setting
    where both disciplines produce key. Bottom: mean link occupancy
    against coherence time; the sequential curves separate at high
    coherence, while the synchronized curves overlap.}
  \label{fig:bottleneck}
\end{figure*}

Figure~\ref{fig:bottleneck} (top) shows the secret-key rate. At the
three shortest coherence settings, sequential produces no key for any
slow-link position, while synchronized produces about $2.7$~bits/s.
At most longer coherence settings, moving the slow link changes the
rate by less than $1\%$ within either discipline. The main exception
occurs at the first setting where sequential produces key,
$\Text=12{,}500~\mu$s. Its rate increases from about $188$~bits/s
with the slow link at hop~$1$ to about $203$~bits/s at hop~$4$,
a spread of roughly $8\%$. At the longest coherence time, both
disciplines reach about $470$~bits/s regardless of slow-link position.

The relative performance of the disciplines also differs from the
homogeneous case. At $\Text=12{,}500~\mu$s, sequential leads by
about $6\%$ with the slow link at hop~$1$ and $14\%$ at hop~$4$.
At every longer coherence setting, synchronized instead has the
higher key rate for all four positions, although its advantage
remains at or below $2.0\%$. From $\Text=125{,}000~\mu$s onward,
both disciplines deliver nearly the same number of end-to-end
pairs, about $3.74\times10^{3}$ per second, while sequential
has slightly lower end-to-end fidelity. For example, at
$\Text=125{,}000~\mu$s, the fidelities are $0.8413$ and
$0.8415$ for sequential and synchronized, respectively.

Figure~\ref{fig:bottleneck} (bottom) shows a stronger dependence
on slow-link position in mean link-buffer occupancy. Under
sequential, the curves coincide at short coherence and begin
to separate above $\Text\approx10^{4}~\mu$s. Occupancy is
highest when the slow link is at hop~$1$ and decreases as it
moves toward hop~$4$. At the longest coherence time, the mean
occupancies are approximately $11$, $7.4$, $4.2$, and $1.1$
for slow-link positions $1$ through $4$. Under synchronized,
the curves remain nearly identical throughout the sweep,
reaching about $11$ at the longest coherence time. These are
the largest mean link-buffer occupancies observed in our
simulations.

Thus, away from the transition region, slow-link position has
little effect on end-to-end key rate under either discipline,
but substantially affects elementary-link-buffer occupancy
under sequential swapping.

\section{Discussion}
\label{sec:discussion}

Because both disciplines use the same trained link controllers at each
setting, their differences arise from how delivered entanglement is stored
and swapped. The homogeneous-chain results show three regimes: short
coherence, where only synchronized produces key; the onset of sequential
operation; and long coherence, where the two rates nearly coincide.
These regimes reflect the different timing requirements of the two
swapping disciplines.

\paragraph{Short-memory regime.}
Synchronized needs all link buffers to contain usable pairs at the same
time, whereas sequential can preserve partial progress between cycles.
In our sequential implementation, however, a newly started chain cannot
be extended until the following cycle. It therefore requires a delivered
pair to remain usable for at least one cycle.

When $\tcut<\tick$, that condition fails. Sequential cannot complete a
chain, while synchronized can still succeed when all $n$ links deliver
within the same cycle, because newly delivered pairs can be used
immediately at zero storage age.

This synchronized pair-delivery rate can be predicted from link-layer
statistics without fitting. Approximating the per-cycle delivery
probability of each link by
$q=1/\langle L_{\mathrm{eps}}\rangle$ and assuming independent links,
the probability that all $n$ links deliver in the same cycle is $q^n$.
The resulting rate is
\begin{equation}
  R_{\mathrm{floor}}
  =
  \frac{q^n}{\tick}
  =
  \frac{1}
       {\langle L_{\mathrm{eps}}\rangle^{\,n}\tick}.
  \label{eq:floorrate}
\end{equation}
For the five shortest-coherence controllers,
$\langle L_{\mathrm{eps}}\rangle=5.11$--$5.19$ cycles at $L=5$~km and
$7.07$--$7.21$ cycles at $L=10$~km. Equation~(\ref{eq:floorrate})
therefore predicts $55$--$59$ and $7.4$--$8.0$ end-to-end pairs/s,
respectively, compared with measured values of $58.2$ and $7.82$.
Using the measured link-buffer occupancies, $q=0.195$ and $0.141$,
gives the same prediction to within $1\%$.

The synchronized pairs in this regime are formed from freshly delivered
elementary pairs and retain $\Fetoe\approx0.844$. Applying the
six-state key fraction gives the small nonzero key-rate floor in
Fig.~\ref{fig:crossover}.

\paragraph{When sequential becomes viable.}
Sequential first becomes possible when a delivered elementary pair can
remain above its storage threshold for one cycle. Setting
$\tcut=\tick$ in Eq.~(\ref{eq:tcut}) gives
\begin{equation}
  \left(\frac{\Tc}{\tick}\right)^{\!*}
  =
  \frac{2}
       {\ln\bigl(x_{\mathrm{del}}/x_{\min}^{1/n}\bigr)}
  \approx 142,
  \label{eq:onset}
\end{equation}
where $x_{\mathrm{del}}=(4\Fdel-1)/3$ and the numerical value uses
$\Fdel\approx0.9575$. A finite positive threshold exists only when
the delivered fidelity exceeds the per-link requirement.

For the observed delivered fidelities, $0.9567$--$0.9575$,
Eq.~(\ref{eq:onset}) predicts thresholds of approximately $142$--$155$.
This agrees with the simulations: sequential produces no key at
$\Tc/\tick=100$ and does at $\Tc/\tick=250$ for both link lengths.
The corresponding storage lifetimes are less than one cycle at the
five settings where sequential fails and reach about $1.8$ cycles
at $L=5$~km and $1.6$ cycles at $L=10$~km at the first setting
where it succeeds.

Equation~(\ref{eq:onset}) is a viability condition for this sequential
implementation, not a general crossover between the two disciplines.
Surviving one cycle is necessary for sequential operation but does not
guarantee a particular rate or an advantage over synchronized. The
estimate also uses the mean delivered fidelity; variation in delivered
fidelity can broaden the onset region.

\paragraph{Long-coherence behavior.}
Once pairs survive across cycles, sequential can preserve partial chains.
In homogeneous chains, this increases its delivery rate relative to
synchronized, with the largest advantage near the first viable setting.
At $L=5$~km, for example, sequential delivers lower-fidelity end-to-end
pairs near this point ($\Fetoe\approx0.829$ versus $0.835$) but still
achieves the higher secret-key rate because it delivers them more often.

As coherence increases further, both disciplines approach the same
link-limited delivery rate. Every end-to-end pair consumes one pair from
each link, so for identical links the delivery rate cannot exceed
$1/(\langle L_{\mathrm{eps}}\rangle\tick)$. At the longest coherence
times, this ceiling is about $7.8\times10^3$ pairs/s at $L=5$~km and
$3.8\times10^3$ pairs/s at $L=10$~km. Sequential reaches about $95\%$
of this limit and synchronized about $94\%$. Their similar delivery
rates and end-to-end fidelities account for the key-rate difference
falling below $1\%$.

\paragraph{Memory use.}
The link-buffer occupancies in Fig.~\ref{fig:occupancy} do not measure
total memory use because sequential also stores partial chains. At
$L=5$~km and the longest coherence time, each completed end-to-end pair
spends about $364~\mu$s in partial-chain storage. With a delivery rate
near $7.4\times10^3$ pairs/s, this corresponds to an average contribution
of about $2.7$ stored partial chains associated with successful
end-to-end deliveries.

Adding this contribution to the link-buffer occupancy summed across
the four links gives at least about $9.7$ stored objects for sequential,
compared with about $12.4$ elementary pairs for synchronized. The
sequential estimate excludes partial chains that expire before completion,
and neither estimate captures peak occupancy. Thus, this comparison does
not establish which discipline requires less total memory. In particular,
the separation between the link-buffer curves should not be interpreted
as the ratio of their total memory requirements.

\paragraph{Effect of a slow link.}
The heterogeneous chains separate rate effects from storage-location
effects. Away from the transition region, moving the $10$~km link changes
the key rate by less than $1\%$ within either discipline. The slowest link
limits the supply of complete sets of elementary pairs, so its position
has little effect on the rate for the chains studied here.

The presence of the slow link also changes which discipline performs
better. At longer coherence times, both deliver end-to-end pairs at
nearly the same rate, but sequential partial chains can wait and decay
at the bottleneck. Its slightly lower end-to-end fidelity is then enough
for synchronized to lead by up to $2\%$. In the homogeneous chain, no
single link systematically delays the partial chains, so preserving
partial progress instead provides a delivery-rate advantage.

The bottleneck position strongly affects where sequential stores
entanglement. When the slow link is first, pairs generated on later links
accumulate in their link buffers. Moving the slow link toward the end
allows more of those pairs to be incorporated into partial chains,
shifting storage away from the link buffers. Synchronized treats the
links symmetrically, so moving the slow link changes which individual
buffer stores pairs but has little effect on the average occupancy.

\paragraph{Longer chains.}
Increasing the number of hops makes both timing requirements more
demanding. In the short-memory regime, synchronized succeeds only when
all links deliver together, so for identical links its rate scales as
$q^n/\tick$ and decreases exponentially with $n$ when $q<1$.

Sequential faces a different limitation. The equal-share per-link
requirement $\Freq(n)$ increases with chain length, so the coherence
needed for one-cycle survival also increases. As the delivered fidelity
approaches $\Freq(n)$ from above, the threshold in
Eq.~(\ref{eq:onset}) diverges. At or below that requirement, increasing
memory lifetime alone cannot provide positive storage time while
satisfying the allocated fidelity threshold.

The controllers used here already reach this limit at $n=5$, where
$\Freq(5)\approx0.9578$ exceeds the largest mean delivered fidelity,
$0.9575$. Five fresh pairs at this fidelity would give an end-to-end
fidelity of about $0.810$, below $\Fmin$. Extending the study beyond four
hops would therefore require retraining the link controllers for the
higher fidelity requirement, which would in turn change their delivery
rates and fidelities. The present results identify these scaling trends
but do not determine which swapping discipline would perform better on
longer chains.

\section{Conclusion and Outlook}
\label{sec:conclusion}

We compared sequential (swap-and-wait) and synchronized (wait-and-swap)
entanglement swapping on a four-hop repeater chain for six-state QKD,
using the same reinforcement-learned link controllers under both
disciplines. Our results show that preserving partial progress is not
uniformly beneficial. When stored pairs cannot remain usable for one
network time step, our sequential implementation produces no key, while
synchronized can still succeed when all links deliver together. Once
sequential becomes viable, it achieves higher key rates in homogeneous
chains, although its advantage decreases as memory lifetime increases.
With one slower link, however, partial chains can wait and decohere at
the bottleneck, allowing synchronized to perform slightly better at
longer memory lifetimes. Moving the slow link has little effect on key
rate away from the transition region but strongly affects sequential
link-buffer occupancy.

These differences can be understood through the timing and fidelity
requirements of the two disciplines. We predict the synchronized
short-memory pair-delivery rate from link-delivery statistics and derive
a closed-form condition for when a delivered pair can survive one network
time step, as required by our sequential implementation. Together, these
analytical results connect the observed operating regimes to the
delivery rate and fidelity of the elementary links, as well as the
available memory lifetime.

The comparison also points toward several extensions. Intermediate
strategies, such as partial synchronization or local
swap-as-soon-as-possible rules, could outperform the two fixed
disciplines studied here. Longer chains require renewed link-layer
optimization to meet higher fidelity requirements, while practical
memory provisioning requires peak occupancy and overflow statistics
beyond the mean link-buffer occupancies reported here. Incorporating
imperfect operations, finite classical-communication delays and gate
times, and multiple simultaneous sessions would further test these
trade-offs under more realistic network conditions. The present work
provides a controlled baseline for developing adaptive network-layer
policies that respond to link quality, memory constraints, and network
traffic.

%% ---- Code availability -------------------------------------------

\section*{Code Availability}
The simulation code, reinforcement-learning controllers, and scripts
used to generate the results presented in this paper are available
in our \href{https://github.com/Pitt-Q-N-Lab-Work/Sequential-vs.-Simultaneous-Entanglement-Swapping-under-Optimal-Link-Layer-Control}{GitHub repository}.

%% ---- Ethics statement --------------------------------------------
\section*{Ethics and Privacy Statement}
This work is a simulation study of entanglement-distribution protocols
and involves no human subjects, personal data, or operational
measurements. We see no direct societal risk beyond the general dual-use
character of quantum-network research; the methods are intended to inform
the design of secure quantum communication infrastructure.

%% ---- Acknowledgments ---------------------------------------------

\begin{acks}
KPS thanks the U.S. Department of Energy, Office of Science, Advanced
Scientific Computing Research (ASCR) program, for support under Award
Number DE-SC0026264. KPS, PK, ASW, and ST thank the PQI Community
Collaboration Awards. KPS thanks Don Towsley for insightful discussions.
\end{acks}

%% ---- Bibliography ------------------------------------------------
\bibliographystyle{ACM-Reference-Format}
\bibliography{references}

\end{document}